\documentclass[11pt,a4]{article}
\usepackage{graphicx}
\usepackage{amsmath}
\usepackage[a4paper, total={6in, 8in}]{geometry}

\title{\bf Accentuating the Competency of 3d Transition Metal Doped $\bf Mg_4$ Clusters towards Molecular Hydrogen Adsorption: A DFT Study}
\author{ Bishwajit Boruah, Bulumoni Kalita\footnote{Corresponding Author, bulumonikalita@dibru.ac.in}\\
Dept. of Physics, Dibrugarh University,Assam, India, 786001}

\begin{document}
\maketitle

\begin{abstract}
Density functional theory $(DFT)$ studies show that doping of $3d TM$ atoms into $Mg_4^{0,+} (TMMg_3^{0,+})$ can alter the endothermic nature of molecular hydrogen adsorption on bare $Mg_4^{0,+}$.  $H_2$ adsorption on $TMMg_3^{0,+}$ clusters depends on the $TM(3d)$ orbital contribution to the frontier molecular orbitals of the clusters.  In $H_2TMMg_3^{0,+}$ complexes, $H_2$ is adsorbed through donation and back donation of electronic charges with $TMMg_3^{0,+}$ clusters.  $H_2$ binding energy is the maximum for $NiMg_3$ (-0.76 eV) and $FeMg_3^+$ (-0.48 eV) among the neutral and cationic $TMMg_3$ clusters, respectively.  The geometry of $H_2$ adsorption complex entirely depends on the geometry of the host cluster.  The lowest energy symmetrical structures of both $NiMg_3$ and $FeMg_3^+$ clusters are less efficient than some slightly higher energy low symmetrical geometrical isomers for adsorption of multiple hydrogen molecules (13.28 $H_2$ wt\% for $NiMg_3$ and 7.89 $H_2$ wt\% for $FeMg_3^+$).  Greater exposer of $TM^{0,+}$ along with its lower Mg coordination number make host clusters more suitable for hydrogen storage.
\end{abstract}

\section{Introduction}\label{sec1}

The present world economy is mostly dependent on the traditional fossil fuels, which take millions of years to form and also releases highly toxic elements into the atmosphere.  In such enormously growing demand of energy, hydrogen is possibly the best alternative with its exceptionally impressive mass-energy density and cleanliness.  However, to commercialize hydrogen as an energy carrier, its storage medium must fulfil the required gravimetric and volumetric density and fast kinetics set by the Department of Energy (DOE), USA  \cite{bib1}. Accordingly, the necessary binding energy of hydrogen to the storage material should be within the range of 0.1eV- 0.8 eV, which lies in between physisorption and chemisorption  \cite{bib2, bib3}. Dissociative chemisorption of hydrogen molecule results in strong metal-hydrogen atom bonds increasing the desorption temperature.  This limits the practical application of hydrogen storage.  This issue can be resolved through adsorption of hydrogen in its molecular form so that its dissociation on the host material resulting in high chemisorption energy can be avoided.  Again, according to the DOE target, the optimum hydrogen storage capacity should be 6.5 wt\% for commercial applications \cite{bib1}.  In order to achieve high hydrogen storage capacity, the host material should possess high gravimetric density of hydrogen.  This parameter is directly proportional to the number of adsorbed $H_2$ molecules and inversely proportional to the molecular mass of the host.  Therefore, the light elements such as Be, Na, Mg, Al, K and Li due to their reasonably less molecular masses are given importance to achieve high gravimetric density of $H_2$ for on-board applications \cite{bib4,bib5,bib6,bib7,bib8}.In a previous study on aromatic bimetallic clusters via density functional theory (DFT) calculation, K. Srinivasu and co-workers have reported high gravimetric density of hydrogen in hydrogenated $Al_4M_2, Be_3M_2, Mg_3M_2$ etc (M=Li, Na, K) clusters and proposed the formation of complexes $Al_4M_2(H_2)_{2n}$ and $Be_3M_2(H_2)_{2n}$ by $Al_4M_2$ and $Be_3M_2$ clusters through adsorption of multiple hydrogen molecules \cite{bib7}. Recently Mg based materials have been widely studied to design $H_2$ storage medium because of their many advantageous properties such as cost effectiveness, abundance, non-toxicity and storage capacity \cite{bib9,bib10,bib11,bib12,bib13,bib14,bib15,bib16,bib17,bib18,bib19}. Yartys et al., in 2019 have reported a very comprehensive review on Mg based hydrogen storage materials including the latest activities, historic overviews and also projecting outlines on future developments putting special importance on improving kinetics and thermodynamic properties \cite{bib12}. Several theoretical and experimental studies reported that the thermodynamic and kinetics of hydrogen adsorption and desorption can be improved by reducing the size of Mg particles to nanoscale \cite{bib15,bib20,bib21,bib22}. Jeon et al., in 2011 have been able to adsorb and release hydrogen on Mg nanoparticles below 200 ºC \cite{bib23}. DFT calculations have shown that hydrogen adsorption properties of Mg clusters are tuneable via varying size, charge state, geometry, dopant element, alloying etc \cite{bib24,bib25,bib26}. ].  Effect of transition metal (TM) doping on hydrogen adsorption properties of different host materials is also an interesting area of research \cite{bib12,bib14,bib27,bib28,bib29,bib30,bib31,bib32,bib33,bib34}. Kubas reported that interaction of $H_2$ with transition metals takes place via donation of electron from $H_2 \sigma$ -orbital to TM vacant d-orbitals followed by back-donation from TM d-orbitals to $H_2 \sigma$*-orbital, which is famously known as Kubas interaction \cite{bib35}. However, binding of $H_2$ varies with different dopant TM as well as with coordination number of the dopant TM in the cluster \cite{bib12,bib14,bib27,bib28,bib29,bib30,bib31,bib32,bib33,bib34,bib35,bib36,bib37,bib38,bib39,bib40}. Jia et al. highlighted the effect of size dependent Rh+ coordination number in $Al_nRh^+$ (n=1-12) clusters on hydrogen molecule adsorption.  It was reported that lower coordination favours stronger hydrogen binding.  Several earlier studies have shown that the catalytic activity of transition metals can greatly improve the hydrogen storage properties of $MgH_2$ \cite{bib14,bib41,bib42}. Xie et al. investigated the hydrogen adsorption properties of MgH2 nanoparticles doped with Ni nanoparticles and reported that the catalytic effect of Ni can be increased by decreasing the size of the nanoparticle \cite{bib41}. In another study Cui et al. have explored the hydrogen storage properties of a series of Mg-TM (TM=Ti, Nb, V, Co, Mo, Ni) nanocomposites and ranked the dehydrogenation performance as $Mg-Ti > Mg-Nb> Mg-Ni> Mg-V> Mg-Co> Mg-Mo$ \cite{bib42}. Charkin and Maltsev have recently studied the reaction kinetics for the hydrogenation of different 3d transition metal (M) doped $Mg_{17}M$ clusters  \cite{bib43,bib44,bib45}. They have reported that Ni shows the most favourable reaction kinetics in the formation of $Mg_{17}MH_2$ systems \cite{bib45}. Trivedi and Bandopadhyay studied the hydrogen adsorption on Rh and Co doped Mg clusters in two different DFT studies and reported $Mg_5Co$ and $Mg_9Rh$ as the most effective hydrogen storage clusters \cite{bib29,bib31}. Ma et al. in another report mentioned that addition of Ti and Nb into $Mg_{55}$ cluster enhances its stability and improve hydrogenation kinetics \cite{bib33}.

 In spite of having the advantage of high gravimetric densities, the study on hydrogen adsorption in small clusters of light alkali and alkali-earth metals is highly limited.  Next, study on such clusters adsorbing hydrogen in molecular form is even scarcer.  However, such studies would be very much useful in designing hydrogen storage materials for practical applications.  Motivated by this fact, we intend to make a comprehensive study on the effect of 3d TM doping in $Mg_4$ cluster towards molecular hydrogen adsorption.  In the present study, we have carried out DFT calculations to explore the competency of gas phase 3d TM doped Mg clusters towards molecular hydrogen storage.  In this regard, we have investigated various factors governing the process.  Accordingly, we have studied the effect of charge states of $TMMg_3$ clusters on molecular hydrogen adsorption; the interplay between the electronic and geometric structures of neutral and cationic $TMMg_3$ clusters as well as their roles in molecular hydrogen adsorption mechanism; nature of adsorption of single and multiple hydrogen molecules on $TMMg_3^{0,+}$ clusters.  The results of this study will guide further investigations on molecular hydrogen adsorption properties of small light metal clusters as well as to tune them for completing the search for hydrogen storage materials.

\section{Methodology}\label{sec2}
All calculations have been carried out in the Gaussian 09 package \cite{bib46} using DFT method.  Mg clusters include features of van der Waals interactions and therefore it is important to include dispersion correction in the Mg cluster calculations in a proper way \cite{bib47}.  Moreover, similar treatment is also necessary for studying the physisorption of molecular H2 in these clusters due to the van der Waals interactions \cite{bib25}  In a recent article, we have established the reliability of dispersion corrected $\omega$B97X-D functional for accurate determination of structure and energetics of gas phase small Mgn clusters \cite{bib47}.  $\omega$97X-D is a hybrid range-separated functional and capable of capturing both short-range and long-range interactions \cite{bib48}. The $\omega$B97X-D method is also reported to be effective in the study of transition metal compounds \cite{bib49}.  Therefore in the present study, we have used $\omega$B97X-D functional for studying the $H_2Mg_4^{0,+}$ and $(H_2)_nTMMg_3^{0,+}$ complexes formed via adsorption of single and multiple hydrogen molecules on $Mg_4^{0,+}$ and $TMMg_3^{0,+}$ clusters, respectively.  6-311G(d,p) basis set is chosen for Mg and H and LANL2DZ basis set is used for TM atoms.  
The binding energy per $H_2$ molecule $(E_b)$ is calculated using the following formula:

\begin{equation}
\label{simple_equation}
E_b= \frac{E(H_2TMMg_3^{0,+})-E(TMMg_3^{0,+})-nE(H_2)}{n}
\end{equation}
\\
The capacity of adsorption of multiple hydrogen molecules for the studied clusters is calculated in terms of wt\% as given by the following formula:

\begin{equation}
\label{simple_equation}
wt\%= \frac{nm(H_2)}{m(TM)+m(3Mg)+m(nH_2)}
\end{equation}
\\Here, E, m and n represents total energy of the corresponding complex/cluster, molecular mass and number of corresponding hydrogen molecules.

\section{Results and Discussions}\label{sec3}
\subsection{Adsorption of single $\bf H_2$ molecule on $\bf Mg_4^{0,+}$ and $\bf TMMg_3^{0,+}$  clusters}

I n a recent study, we have shown that the electronic properties of  $TMMg_3^{0,+}$  clusters are mainly governed by $TM^{0,+} (3d)$ orbitals \cite {bib47}. Therefore, in the present study, we have chosen the  $TMMg_3^{0,+}$ clusters along with $Mg_4^{0,+}$ to host the adsorption of molecular hydrogen. Accordingly, we have performed symmetry unrestricted full geometry optimizations via DFT calculations for all possible $H_2$ adsorption geometries of $Mg_4^{0,+}$ and $TMMg_3^{0,+}$ clusters in different spin multiplicities. It is to be noted that $H_2$ adsorption on bare $Mg_4^{0,+}$ clusters is an endothermic process (Table A1 in supplementary file (SF)). A previous study with a different computational level has also reported very weak exothermic adsorption of $H_2$ on $Mg_4$ cluster \cite{bib24}. Among the 3d TM series, only doping of V and Ni into $Mg_4$ could alter its nature of binding with $H_2$ as shown by the binding energy values of -0.03 eV and -0.76 eV, respectively (Table A1). Therefore, the lowest energy structures of only these two neutral hydrogen adsorption complexes ($H_2VMg_3$ and $H_2NiMg_3$) are shown in Figure 1. Increment of binding energy of $H_2$ after doping of transition metal atoms has also been reported earlier [33]. On the other hand, $H_2$ binding has been observed to be exothermic in all the doped cationic clusters. The binding energy values of the cationic complexes are presented in Table A2 and they are found to be in the range of -0.01 eV to -0.48 eV. $H_2FeMg_3^{0,+}$ and $H_2VMg_3^{0,+}$ complexes possess the highest and the lowest binding energy of $H_2$, respectively. The lowest energy structures of all the $H_2TMMg_3^+$ complexes are also shown in Figure 1. The spin multiplicities of all the lowest energy structures are presented in Figure 1. Structures of all other neutral and cationic complexes are presented in Figure A1 for reference. Almost all the lowest energy $H_2TMMg_3^{0,+}$ complexes carry the structural feature of $H_2$ being adsorbed on top of the TM, except for $ZnMg_3^+$ (Figure 1). The bond lengths of the stable neutral and cationic clusters are mentioned next to their geometries in Figure 1 (also shown in Table A1 and A2). The $TM-H_2$ bonds in the lowest energy $H_2TMMg_3^{0,+}$ complexes are significantly lower than the $Mg-H_2$ bond in $H_2Mg_4$ (Table A1). This indicates enhancement in interaction of the hydrogen molecule with the Mg tetramer due to TM doping. The $H_2NiMg_3$ complex is found to exhibit the shortest $TM-H_2$ bond length supporting its highest $H_2$ binding energy value. In $H_2NiMg_3$, H-H bond length elongates to 0.84 \AA but the average Ni-Mg bond length gets shortened to 2.39 \AA from that of 2.61 \AA in bare $NiMg_3$ cluster \cite{bib47}. The $TM^+-H_2$ bond length values in $H_2TMMg_3^+$ complexes are in the range of 1.84 \AA – 3.27 \AA and $Zn^+-H_2$ is found to be the longest bond. The H-H and TM-Mg bond lengths in all the $H_2TMMg_3^+$ complexes are found to be in the range of 0.75 \AA – 0.78 \AA, 2.58 \AA – 3.1 \AA, respectively.

\begin{figure}
\centering
\includegraphics[width=0.8\textwidth]{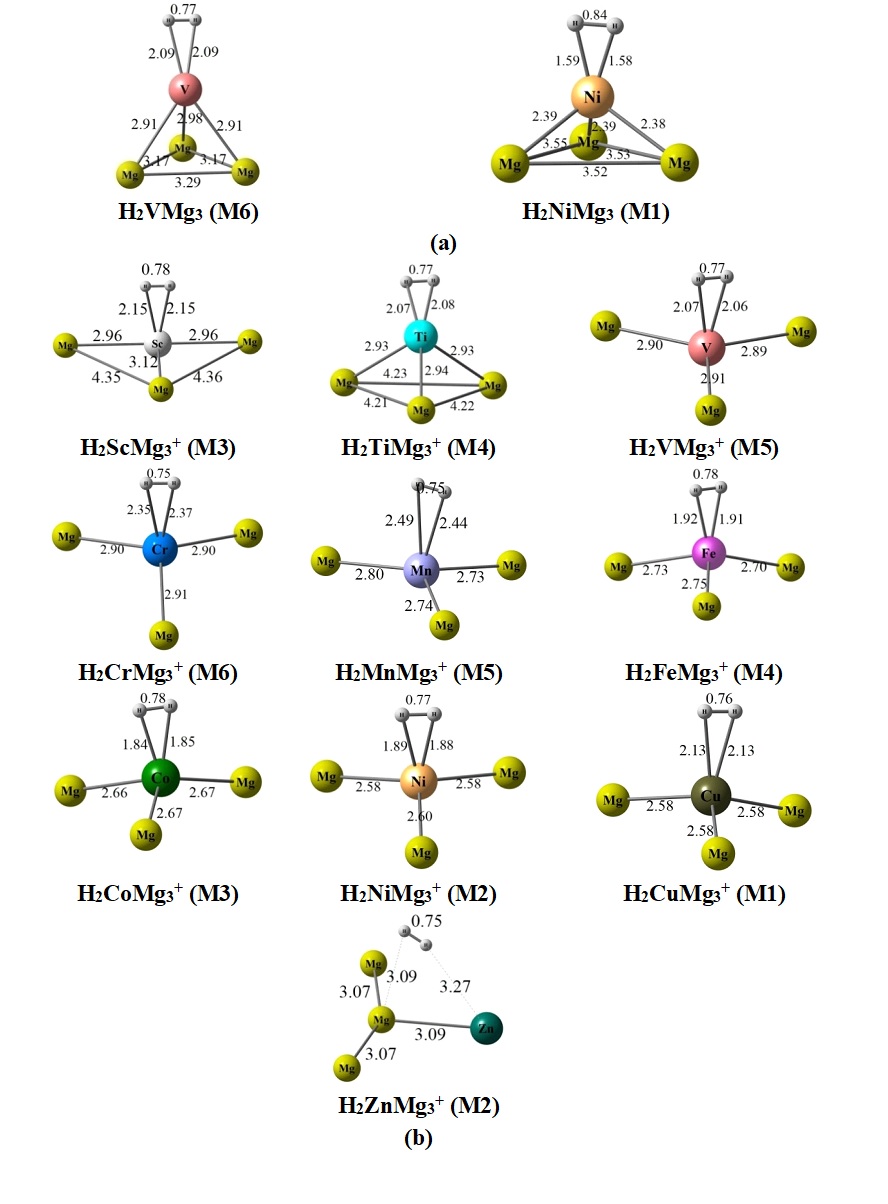}
\caption{\bf Molecular hydrogen adsorbed (a) $\bf H_2TMMg_3$ (b) $\bf H_2TMMg_3^+$ complexes (bond lengths are in \AA). M represents spin multiplicity.}\label{fig1}
\end{figure}

\newpage
\subsubsection{Electronic structures of $\bf TMMg_3^{0,+}$  clusters and adsorption of single $\bf H_2$ molecule}
To explore the nature of interaction of a hydrogen molecule with $TMMg_3^{0,+}$ clusters, we have calculated the percentage of TM(3d) contribution to the HOMO of $TMMg_3^{0,+}$ clusters.  These results are based on molecular orbital composition calculations available in Multiwfn software \cite{bib50}.  It is important to mention here that in this calculation, singlet spin state of $NiMg_3$ has only been considered although there is possibility of getting a triplet state too \cite{bib47}.  These two spin singlet and triplet isomers of $NiMg_3$ are almost equally stable with energy difference of only 0.01 eV.  The triplet isomer is not found to have any TM(3d) contribution to its HOMO.  Moreover, since the most stable $H_2NiMg_3$ complex exists in singlet spin state (section 3.1) and also the other $TMMg_3^{0,+}$ clusters retain their spin states in their corresponding $H_2TMMg_3^{0,+}$ complexes, we have considered the singlet state electronic structure of $NiMg_3$ cluster to study molecular hydrogen interaction.  This validity of this argument is justified by the range of $H_2$ binding energies that is reasonably smaller than the chemisorption process (section 3.1) and hence it is unlikely that the spin state of the cluster may get changed after formation of $H_2TMMg_3^{0,+}$ complexes.  Further analyses to be presented in the following discussion will provide more insight into this aspect.   
We have plotted the calculated values of percentage contribution of TM(3d) towards the HOMO of $TMMg_3^{0,+}$ clusters with the $H_2$ binding energies (BE) in the $H_2TMMg_3^{0,+}$ complexes and have presented in Figure 2.  This figure clearly indicates that the binding energy of molecular hydrogen in $H_2TMMg_3^{0,+}$ complexes is a direct consequence of the TM(3d) orbital contribution to the HOMOs of  the host $TMMg_3^{0,+}$ clusters.  This is due to the fact that presence of TM(3d) orbitals in the HOMOs of the host cluster favours $H_2$ binding with the TM atoms of the cluster via Kubas interaction \cite{bib35}.  Figure 2(a) shows that except for $VMg_3$ and $NiMg_3$, the TM(3d) contribution in HOMO of all the other clusters is absent making molecular hydrogen adsorption unfavourable in these clusters. Interestingly, electron deficiency in the cationic clusters pushes their HOMOs down in energy with reference to the neutral clusters, which make them capable of getting contributed from the TM(3d) orbitals.  This shifting of HOMO levels in $TMMg_3^+$ clusters was already shown in the density of states (DOS) analysis in our previous study \cite{bib47}.  $ZnMg_3^+$, although an exception in this regard, exhibits exothermic nature of binding with $H_2$.  However, $H_2$ binding energy with $ZnMg_3^+$ is the lowest among the cationic clusters, which supports the largest bond length value of $Zn^+-H_2$ as discussed above.  

\begin{figure}
\centering
\includegraphics[width=0.6\textwidth]{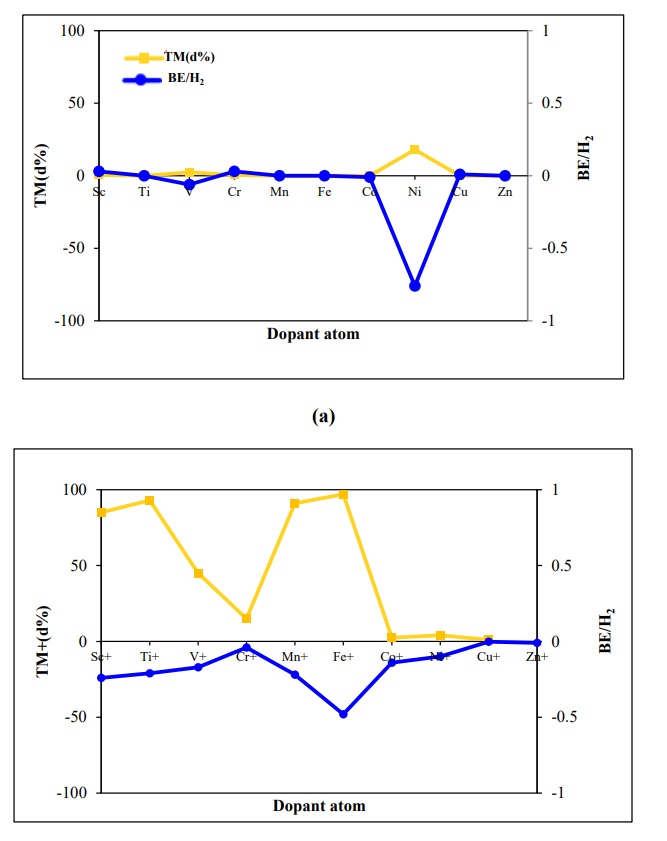}
\caption{\bf Molecular hydrogen adsorbed (a) $\bf H_2TMMg_3$ (b) $\bf H_2TMMg_3^+$ complexes (bond lengths are in \AA). M represents spin multiplicity.}\label{fig1}
\end{figure}

\subsubsection{Adsorption mechanism of single $\bf H_2$ molecule on $\bf NiMg_3$ cluster}
To study the interaction mechanism of $H_2$ with the host clusters, we have performed natural bond orbital (NBO) and density of states (DOS) calculations.  The results of NBO calculations derived from Gaussian 09 have been utilized to interpret the interaction of frontier molecular orbitals of the $H_2$ molecule and the clusters.  The DOS calculations are carried out separately in Multiwfn software.  Since hydrogen binding is found to be the maximum in $H_2NiMg_3$ complex, we are now going to discuss the interaction mechanism of hydrogen molecule with $NiMg_3$ cluster considering the most stable geometry of the complex.  The results for the rest of the neutral and cationic complexes are provided in the SF (Table A3-A6 and Figure A2-A5).  In Figure 3, we have shown a schematic representation of the interaction of the frontier molecular orbitals of $H_2$ and $NiMg_3$ cluster as obtained from the molecular orbital analysis.  According to perturbation theory, the interaction of HOMO (filled orbital) of one system with LUMO (empty orbital) of another can lead to overall stabilization through donor-acceptor bonding interaction\cite{bib3}.  It is observed that the HOMO of $H_2$ having energy -14.10 eV interacts with the LUMO of $NiMg_3$ with energy (-0.83 eV) resulting in more stable bonding orbitals of the $H_2NiMg_3$ complex (-15.42 eV).  Similarly, HOMO (-6.16 eV), HOMO-1 (-6.17 eV) of $NiMg_3$ interact with the LUMO of $H_2$ (-3.78 eV) forming $H_2NiMg_3$ bonding orbitals of energies -6.42 eV and -7.61 eV.  Therefore, interaction of $H_2$ molecule with $NiMg_3$ is mediated via the Ni atom and the interaction takes place due to donation and back donation of electronic charges, i.e., the Kubas interaction \cite{bib35}.  Further, careful observation of the donor-acceptor data of NBO results for $H_2NiMg_3$ (Table 1) reveals the presence of additional interaction in the complex arising from some of the molecular orbitals not included in Figure 3(a).  Significant values of second order energy correction (E(2)) in Table 1 suggests interactions of $H_2$ bonding orbital (BD) with spd hybridised orbital of Ni (LP*) as well as $H_2$ anti-bonding orbital (BD*) with Ni(3d) (LP).  These interactions are also confirmed from the partial density of states (PDOS) analysis shown in Figure 3(b).  The PDOS plots clearly show the presence of interactions between $H_2$ and Ni in $H_2NiMg_3$ at various energy scales.  The corresponding energies are precisely evaluated as -15.42 eV, -8.5 eV, -7.61 eV and 6.42 eV in consultation with the molecular orbital analyses to be provided in section 3.2.1.  Similar type of interactions are also seen in other $H_2TMMg_3^{0,+}$ complexes except for $H_2ZnMg_3^+$ (Figure A2-A3 in SF).  $H_2ZnMg_3^+$ is survived with weak van der Waals type interaction rather than the Kubas interaction.  This observation is consistent with that discussed earlier in the beginning of section 3.1.

\begin{figure}
\centering
\includegraphics[width=0.9\textwidth]{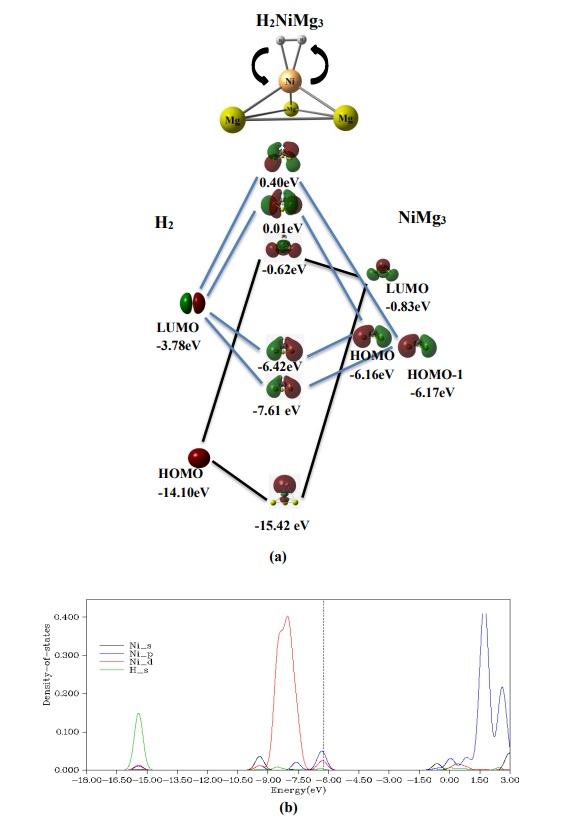}
\caption{\bf: (a) Schematic representation of the interaction of the frontier molecular orbitals of$\bf H_2$ and $\bf NiMg_3$ cluster (b) PDOS plots of $\bf H_2NiMg_3$ complex.}\label{fig3}
\end{figure}
\clearpage

\begin{table}
\begin{center}
\caption{\bf Donor acceptor analysis of $\bf H_2NiMg_3$ complex. E(2) represents the second order energy correction.}\label{<table-label>}%
\begin{tabular}{@{}llll@{}}
           & $\bf H_2NiMg_3$	 &          \\
  \bf Donor & & \bf Acceptor \\
  BD (H – H) & &LP* (Ni)
s (34.48\%) p 1.72 (59.23\%) \\&&d 0.18 (6.29\%),E(2) = 4.52 eV\\
LP (Ni) 
s (2.58\%) \\p 1.43 (3.70\%) d 36.28 (93.72\%)&&BD* (H – H),E(2) = 1.21 eV
\end{tabular}
\end{center}
\end{table}

\subsection{\bf Adsorption of multiple $\bf H_2$ molecules on $\bf TMMg_3^{0,+}$ clusters}
Now, we are going to find the capacity of the $TMMg_3^{0,+}$ clusters in adsorbing multiple $H_2$ molecules, which will be parameterized by the gravimetric density calculations.  The gravimetric density is considered as a key attribute of an efficient hydrogen storage material and it is represented as \%wt (equation number (2)) of hydrogen adsorbed by the material.  Higher the value of \%wt of $H_2$, higher will be the gravimetric density and hence efficiency of a material in adsorbing $H_2$.  In the section 3.1., $NiMg_3$ and $FeMg_3^+$ have been confirmed as the most efficient neutral and cationic $TMMg_3$ clusters for adsorption of single $H_2$ molecule.  Therefore, we are now going to examine the capacity of these clusters in adsorbing multiple $H_2$ molecules within the optimum binding energy range of 0.1 eV – 0.8 eV per $H_2$ molecule as mentioned in the introduction.  

\subsubsection{\bf Adsorption of multiple $\bf H_2$ molecules on $\bf NiMg_3$ cluster}

$NiMg_3$ clusters adsorbing n number of $H_2$ molecules will be represented as $(H_2)_nNiMg_3$ complexes and the corresponding lowest energy optimized structures are shown in Figure 4(a).  First, we have performed the geometry optimization for the $(H_2)_nNiMg3$ complexes through stepwise addition of $H_2$ molecules to the lowest energy tetrahedral structure of $H_2NiMg_3$ shown in Figure 1(a).  The binding energy per $H_2$ molecule (BE/$H_2$) is found to decrease gradually with the number of hydrogen molecules (-0.76 eV -- -0.10 eV).  Ultimately, six numbers of $H_2$ in total could get adsorbed on $NiMg_3$ cluster within the required binding energy range.  It is noticeable in Figure 4(a) that the first $H_2$ molecule maintains its position on-top of $NiMg_3$ even after successive addition of further five numbers of $H_2$.  It is also obvious that the interactions of five out of the six $H_2$ molecules are taking place only in one side of $NiMg_3$, i.e., towards the Ni atom site.  Based on these two factors, it has been realized that the position of the first $H_2$ molecule as well as the compact three-dimensional geometry of the host may be responsible for creating congestion among multiple $H_2$ molecules hindering the process from continuing further.  We have therefore become more curious to know about the effect of geometry of the host cluster on its hydrogen gravimetric density.  Accordingly, we have next considered the planar geometry (P) of $NiMg_3$ cluster, which has already been reported in our previous study as a stable isomer of $NiMg_3$ with only 0.6 eV higher in energy than the most stable tetrahedral (T) $NiMg_3$ isomer [47].  Figure 4(b) shows the optimized structures of $(H_2)_nNiMg_3$ complexes in the $NiMg_3$(P) isomer.  The binding energy variations with number of $H_2$ molecules for the T and P isomers are compared in Figure 5.  Interestingly, BE/$H_2$ in P isomer slightly increases for the second $H_2$ and starts decreasing gradually thereafter.  The factors affecting the distinguishing behaviour of the T and P isomers towards adsorption of multiple $H_2$ molecules will be discussed in the following section.  It is also observed that in the planar geometry, $NiMg_3$ cluster can adsorb up-to ten hydrogen molecules within the optimum binding energy range.  The calculated wt\% values of $H_2$ in the T and P isomers of $NiMg_3$ are found to be 8.42 and 13.28, respectively.  This clearly confirms our intuition that the geometry of the host cluster plays a crucial role in governing the adsorption of multiple hydrogen molecules in small clusters.
\begin{figure}
\centering
\includegraphics[width=0.9\textwidth]{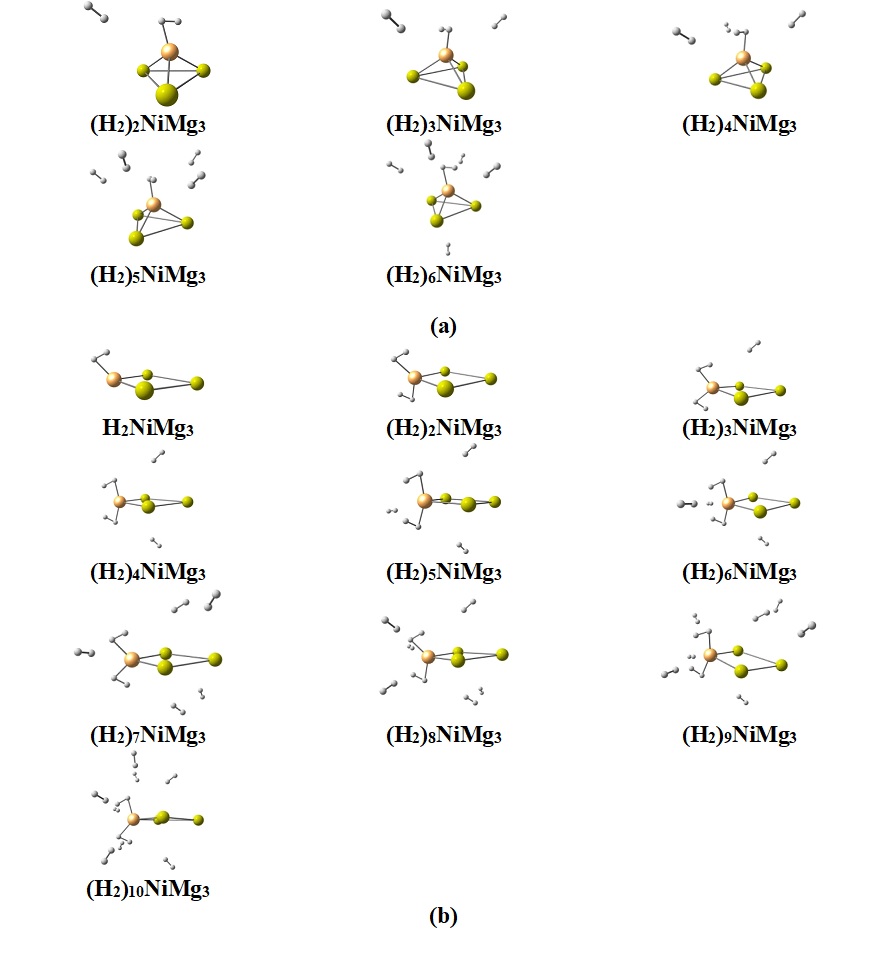}
\caption{\bf Multiple $\bf H_2$ adsorption on (a) tetrahedral (T) and (b) planar (P) isomer of $\bf NiMg_3$ cluster.}\label{fig4}
\end{figure}
\clearpage
\begin{figure}
\centering
\includegraphics[width=0.7\textwidth]{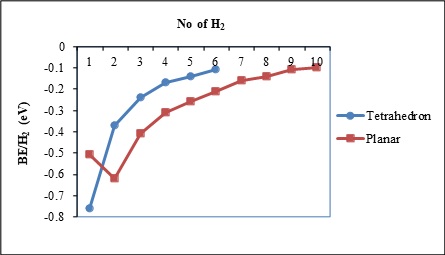}
\caption{\bf: Variation of BE/$\bf H_2$ with no of $\bf H_2$ on tetrahedral and planar isomer of$\bf NiMg_3$ cluster.}\label{fig5}
\end{figure}

\paragraph{\bf 3.2.1.1 Adsorption mechanism of multiple $\bf H_2$ molecules on structural isomers of $\bf NiMg_3$ cluster}
\paragraph{\bf 3.2.1.1.1 Electron density distribution and adsorption of multiple $\bf H_2$ molecules}
\paragraph{}
We have just seen that the planar isomer of $NiMg_3$ can accommodate larger number of $H_2$ molecules than the tetrahedral isomer.  Moreover, there is a distinct jump in $H_2$ binding energy while going from the first to third $H_2$ molecule (Figure 5) in P isomer.  This behavioural difference between T and P isomers of $NiMg_3$ can be explained from careful investigation of their geometrical and electronic structures.   It is observed that the adsorption site for the first $H_2 $ is not similar in the T and P isomers, which affect the adsorption mode of the second $H_2$.  Figure 6 shows the electrostatic potential (ESP) surface plots of these isomers before and after adsorption of $H_2$. The red and blue colours represent the electronegative and electropositive surfaces, respectively.  The upper panel in Figure 6(a) presents the ESP surfaces of T isomer of $NiMg_3$ cluster before $H_2$ adsorption.  Here, the electronegative surfaces are observed around Ni–Mg bonds indicating the presence of electron density.  In the same isomer, the electropositive surfaces are observed on Ni as well as all the Mg atoms.  Due to adsorption of $H_2$ molecule on top of the Ni atom in T isomer, the electron density existing over the Ni–Mg bonds gets delocalized and reduce the electropositive surface on Ni atom as seen from the lower panel of Figure 6(a).  As a result of this, it becomes difficult for the next $H_2$ to reach and interact with Ni atom in T isomer.  Therefore, the second $H_2$ is not capable to alter the position of the first and hence electronic structure of the adsorption complex with only the first $H_2$ molecule is retained in the T isomer.  Because of this, the adsorption configuration and the associated electronic structure of $H_2NiMg_3$ complex become effective in controlling the adsorption geometry as well as strength of interaction for successive hydrogen molecules.  That is why, the binding energies decrease with increasing number of $H_2$ molecules (Figure 5).  In the P isomer, the electronegative surfaces are observed around Ni–Mg and Mg–Mg bonds whereas electropositive surfaces localize over Ni as well as all of the Mg atoms before $H_2$ adsorption (upper panel of Figure 6(b)).  In this isomer, the first $H_2$ is adsorbed on the Ni atom in a tilted geometry, which could provide ample opportunity to the next $H_2$ molecule to approach the Ni atom and interact with it.  The ESP plots for P isomer after the first $H_2$ adsorption (lower panel in Figure 6(b)) shows that the electropositive surface from the top of Ni atom shifts towards its side opposite to the adsorbed $H_2$ molecule.  This new position of the electropositive surface actually facilitates the adsorption of the second $H_2$.  It is important to note here that the second $H_2$ molecule also acquires a tilted geometry making further adsorption of $H_2$ favourable.  This process could therefore ultimately result in adsorption of a reasonably higher number of $H_2$ molecules in $NiMg_3$ cluster in its planar isomer than that in the tetrahedral isomer.
\begin{figure}
\centering
\includegraphics[width=0.9\textwidth]{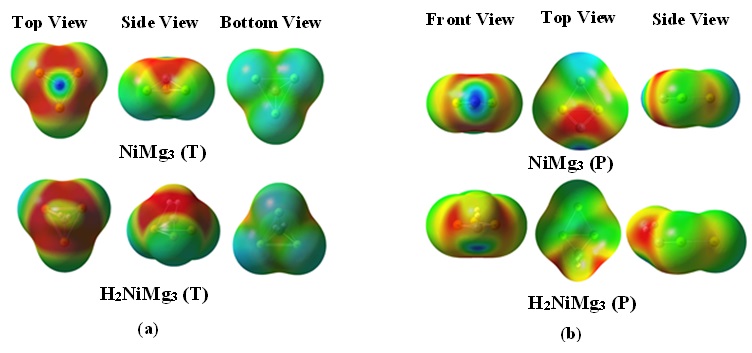}
\caption{\bf Electro static potential (ESP) plots of T and P isomers of $\bf NiMg_3$ cluster before and after adsorption of $\bf H_2$.}\label{fig6}
\end{figure}

\newpage
\paragraph {\bf 3.2.1.1.2. Molecular orbital analysis and adsorption of multiple $\bf H_2$ molecules}
\paragraph{} Figure 7(b) and 7(c) show the molecular orbitals (MOs) of T and P isomers of $NiMg_3$ cluster before and after adsorption of single $H_2$ molecule.  The MOs of bare $Mg_4$ and the $H_2$ molecule are also shown in Figure 7(a) for comparison.  We can see here that the positive and negative iso-surfaces of the frontier MOs in $Mg_4$ are very well separated in contrast with those in $NiMg_3$.  Such separation prohibits the interaction of $Mg_4$ with molecular hydrogen.  However, presence of Ni(3d) orbital contribution in the frontier MOs of $NiMg_3$ isomers (section 3.1.1. and Figures 7(b) and 7(c)) changes their shapes and reduces the separation between the positive and negative iso-surfaces at the Ni atom site.  Due to this, the $H_2$ molecule is captured by the $NiMg_3$ isomers on top of Ni between the two iso-surfaces.  Further, orientations of the MOs in $NiMg_3$ cluster (HOMO and HOMO-1 in T isomer; HOMO, HOMO-1 and HOMO-2 in P isomer) also indicate their possible interactions with the LUMO of H2 molecule.   Such interactions constitute the MOs (HOMO-1, HOMO-2) of $H_2NiMg_3$ complex corresponding to the T isomer of $NiMg_3$ (Figure 7(b)).  In this complex, the HOMO-6 orbital shows mixing of Ni 3dxz orbital with H2 (LUMO).  Similarly, the HOMO-7 and HOMO-8 orbitals of the same complex are due to the interaction of $H_2$ bonding orbital with the LUMO of $NiMg_3$ (T). In P isomer of $NiMg_3$, Ni(3d) orbitals in the xy plane combine with Mg orbitals and forms jellium type orbitals (HOMO, HOMO-1 of $NiMg_3$ (P) in Figure 7(c)), whereas 3d orbitals in xz and yz planes remain isolated (HOMO-2 of $NiMg_3$ (P) in Figure 7(c)).  In the $H_2NiMg_3$ (P) complex, the $H_2$ bonding orbital mixes with the LUMO of $NiMg_3$ (P) and gives rise to HOMO-1, HOMO-7 and HOMO-8 orbitals of the complex (Figure 7 (c)).  On the other hand, $H_2$ (LUMO) interaction with Ni(3dxz) orbital forms the HOMO-6 orbital of $H_2NiMg_3$ (P), whose orientation compels the $H_2$ molecule to get adsorbed only on one side of Ni in the complex helping the next hydrogen to interact from the opposite side.

\begin{figure}
\centering
\includegraphics[width=0.9\textwidth]{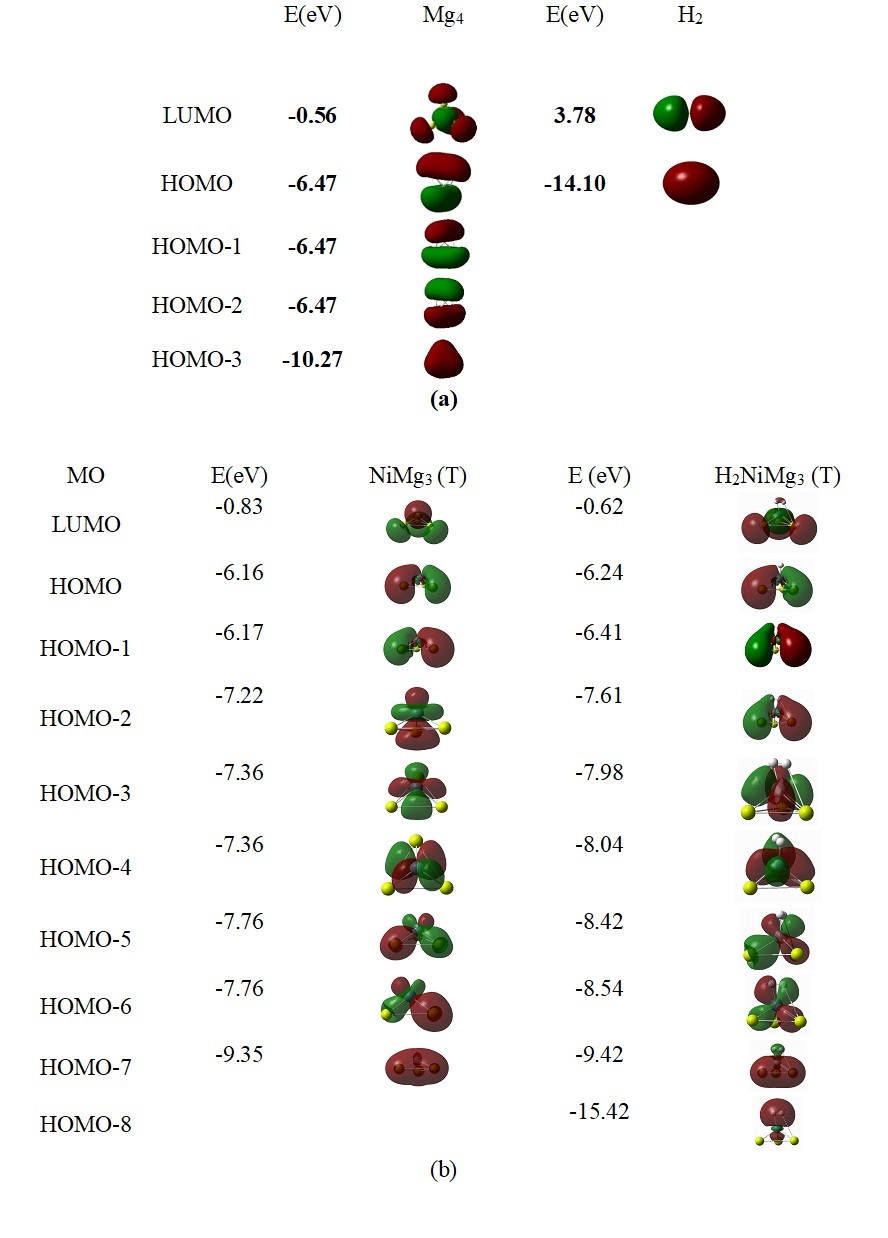}
\end{figure}

\begin{figure}
\centering
\includegraphics[width=0.9\textwidth]{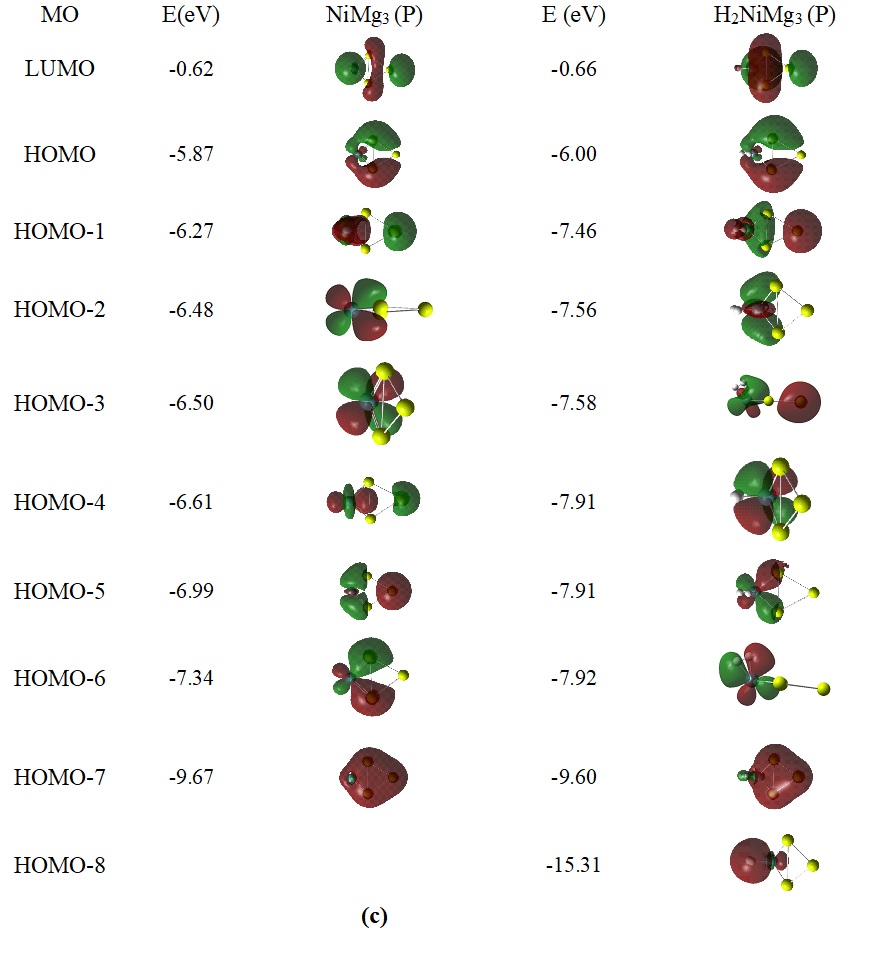} 
\caption{\bf Molecular orbitals of (a) $\bf Mg_4$ and $\bf H_2$ (b) $\bf NiMg_3/H_2NiMg_3$ (T) (c) $\bf NiMg_3/H_2NiMg_3$ (P) complexes.}\label{fig7}
\end{figure}
\clearpage

\paragraph {\bf 3.2.1.1.3. Charge transfer and adsorption of multiple $\bf H_2$ molecules}
\paragraph {}
Table 2 shows the natural electronic configurations of $(H_2)_nNiMg_3$ (n=1-3) complexes along with those of bare $NiMg_3$ isomers.  The corresponding data for the rest of the complexes are provided in the SF (Table A7).  The computed electronic configurations for free Mg and Ni atoms are found to be Mg [core]3s(2.00) and Ni [core]4s(1.25)3d(8.74), respectively.  With reference to these, Ni(4s) orbital in the lowest energy $NiMg_3$ (T) cluster has been observed to donate electronic charge to all the Mg atoms and gain significant electron density in its 4p and 3d orbitals.  The Mg atoms loss electron density from their 3s orbitals and gain in 3p orbitals.  The electronic configurations of all the Mg atoms in the T isomer are identical, which signifies equivalent nature of interaction between each of them and the Ni atom (Table 2).  Charge transfer process in P isomer involves similar set of molecular orbitals of Ni and Mg as that of T isomer.  However, the mechanisms are different in various ways.  The overall gain in electron density by Ni in the P isomer is less compared to that in T isomer.  Unlike T isomer, only the Mg atoms sitting near to Ni in P isomer (Mg1 and Mg2) participate in the charge transfer process.  Moreover, the gain in charge density in the 3p orbitals of the two interacting Mg atoms in P is almost three times higher than that of each Mg atom in isomer T.  The Mg atom away from Ni (Mg4), in exception, is found to have non-equivalent electronic configuration with comparatively less change in its electron density than the other two Mg atoms.  After adsorption of the first $H_2$ molecule, there are much higher charge gain and lower charge loss in Ni(4p) and Ni(4s) orbitals, respectively, in isomer T than those in isomer P (Table 2).  As a result of this, the resultant gain in electron density by Ni atom in the T is more than that in the P isomer.  This property helps in strengthening the interaction of single hydrogen molecule with T isomer of $NiMg_3$ over the P isomer.  The higher electron withdrawal nature of Ni in $H_2NiMg_3$ (T) complex can be attributed to its isotropic Mg environment, where all the three Mg atoms retain their equivalent electron density distributions rendering their equal participations in $H_2$ adsorption.  As expected, the Mg1 and Mg2 atoms in $H_2NiMg_3$ (P) complex carry similar charge distributions but the isolated Mg4 atom behaves in a totally different way.  This also confirms the unequal participations of the Mg atoms in the planar geometry of $H_2NiMg_3$ complex making its interaction with $H_2$ relatively less strong than the tetrahedral geometry.  The donor acceptor interaction for the second $H_2 in (H_2)_2NiMg_3$ (T) is found to be weak resulting in insignificant changes of electronic charges in the corresponding Ni (4p) and Mg(3s) orbitals.  Interestingly, in spite of relatively weaker interaction for the first hydrogen molecule in $H_2NiMg_3$ (P), Ni atom in $(H_2)_2NiMg_3$ (P) complex gains significant electrons in its 4p orbital accompanied by very little electron loss from the 3s orbitals of Mg1 and Mg2.  Therefore, we can say that the suppressed interaction in $H_2NiMg_3$ (P) is favourable for interaction with the second $H_2$ molecule.   But stronger interaction in $H_2NiMg_3$ (T) is responsible for its inert nature towards the next $H_2$ molecule.  Both these cases, however, exhibit weaker donor acceptor interactions for further addition of $H_2$ molecules and hence their BE/$H_2$ values have similar decreasing trends afterwards (Figure 5).

\begin{table}
\begin{center}
\caption{\bf Natural electronic configurations of $\bf NiMg_3$ isomers and their $\bf H_2$ adsorbed complexes (atom numbers are provided for reference).}\label{<table-label>}%
\begin{tabular}{@{}llll@{}}
\includegraphics[width=0.13\textwidth]{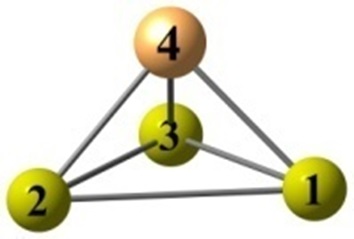} && \includegraphics[width=0.15\textwidth]{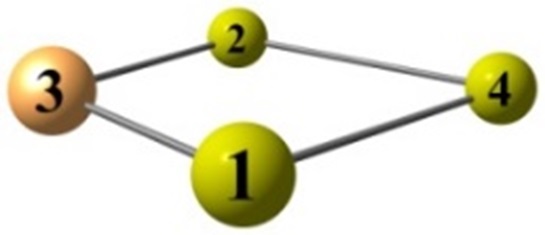} \\
$\bf NiMg_3$ (T) && $\bf NiMg_3$ (P)\\
\includegraphics[width=0.15\textwidth]{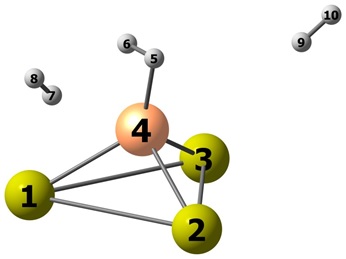} && \includegraphics[width=0.15\textwidth]{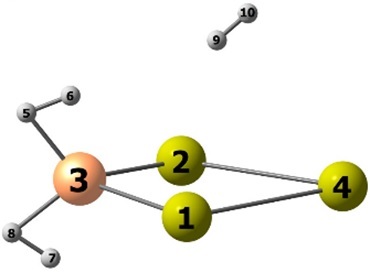} \\
$\bf (H_2)_3NiMg_3$ (T) && $\bf (H_2)_3NiMg_3$ (P)\\
 \bf  T && \bf P\\
& $\bf NiMg_3$ &\\
Mg1 3S( 1.40)3p( 0.13)3d( 0.01)4p( 0.01)&&Mg1 3S(1.44)3p(0.36)3d(0.01)4p(0.01) \\
Mg2 3S( 1.40)3p( 0.13)3d( 0.01)4p( 0.01)&&Mg2 3S(1.44)3p(0.36)3d(0.01)4p(0.01) \\
Mg3 3S( 1.40)3p( 0.13)3d( 0.01)4p( 0.01)&&Ni3 4S(0.65)3d(9.65)4p(0.16)5S(0.01)5p(0.01) \\
Ni4  4S( 0.67)3d( 9.70)4p( 0.98)5p( 0.01)&&Mg4 3S(1.65)3p(0.22)4S(0.01) \\

& $\bf H_2NiMg_3$&\\
Mg1 3S(1.29)3p(0.15)4p(0.01)&& Mg1 3S(1.47)3p(0.28)4p(0.01)\\
Mg2 3S(1.29)3p(0.15)4p(0.01)&& Mg2 3S(1.47)3p(0.28)4p(0.01)\\
Mg3 3S(1.29)3p(0.15)4p(0.01)&& Ni3 4S(0.49)3d(9.59)4p(0.40)5p(0.08)\\
Ni4   4S(0.63)3d(9.62)4p(1.49)5p(0.01) && Mg4 3S(1.74)3p(0.16)\\
H5    1S(0.94)   && H1 1S(1.02)2S(0.01) \\
H6   1S(0.94)   && H2 1S(0.98)2S(0.01)\\

& $\bf (H_2)_2NiMg_3$&\\

Mg1 3S(1.29)3p(0.15)4p(0.01)&&   Mg1 3S(1.39)3p(0.25)4p(0.01)\\
Mg2 3S(1.29)3p(0.15)4p(0.01)&&  Mg2 3S(1.39)3p(0.25)4p(0.01)\\
Mg3 3S(1.29)3p(0.15)4p(0.01)&&  Ni3 4S(0.49)3d(9.54)4p(0.92)\\
Ni4 4S(0.63)3d(9.62)4p(1.51)5p(0.01)&& Mg4 3S(1.75)3p(0.15)\\
 H5 1S(0.94)&& H5 1S(0.96)\\
 H6 1S(0.94)2S(0.01)&& H6 1S(0.95)\\
 H7 1S(1.00)&& H7 1S(0.96)\\
 H8 1S(0.99)&& H8 1S(0.95)\\

& $\bf (H_2)_3NiMg_3$&\\

Mg1 3S(1.28)3p 0.15)4p(0.01)&& Mg1 3S(1.37)3p(0.27)4p(0.01)\\
Mg2  3S(1.26)3p(0.15)4p(0.01)&& Mg2 3S(1.37)3p(0.27)4p(0.01)\\
Mg3  3S(1.29)3p(0.15)4p(0.01)&& Ni3  4S(0.49)3d(9.54)4p(0.92)\\
Ni4  4S(0.63)3d(9.62)4p(1.54)5p(0.01)&& Mg4 3S(1.72)3p(0.17)\\
 H5  1S(0.94)&& H5  1S(0.95)\\
 H6  1S(0.94)2S(0.01)&& H6 1S(0.95) \\
 H7  1S(0.99)&& H7  1S(0.95) \\
 H8  1S(1.00)&& H8 1S(0.96) \\
 H9  1S(0.99)&& H9 1S(1.00)\\
 H10  1S(1.00)&& H10  1S(0.99)\

\end{tabular}
\end{center}
\end{table}

\subsubsection {Multiple $\bf H_2$ adsorption of $\bf FeMg_3^+$ }
Among the cationic clusters, $FeMg_3^+$ possesses the highest binding energy for single $H_2$ adsorption.  Therefore, we have checked the multiple $H_2$ adsorption capacity of this cationic cluster only and accordingly calculated the $H_2$ wt\% for it.  So far, we have seen that the structural geometry of $NiMg_3$ isomers governs the multiple $H_2$ adsorption process for the cluster.  With the same logic, we have decided to proceed with two different structural isomers of $FeMg_3^+$.  In this regard, the most stable isomer of $FeMg_3^+$ (Isomer I) along with another one (Isomer II) with 0.16 eV higher in energy are considered.  In these two isomers, subjection of $Fe^+$ is different as shown in Figure 8.  Calculations reveal that the isomer II can adsorb up to five $H_2$ molecules (7.26 wt\% $H_2$) against maximum of four $H_2$ molecules in isomer I (5.89 wt\% $H_2$) within the optimum binding energy range of 0.1-0.8 eV.  The evolutions of binding energy per $H_2$ molecule for $FeMg_3^+$ isomers are shown in Figure 9.  It is seen here that the relatively higher BE value for the first $H_2$ molecule in Isomer I is not capable of keeping it competent for multiple $H_2$ adsorption.  This is in consistent with that observed for $NiMg_3$ isomers.  Actually, position of $Fe^+$ in Isomer II is more suitable than that in IsomerI for interaction with the $H_2$ adsorbate.  This configuration in Isomer II is maintained even after successive $H_2$ adsorption.  The nature of the electrostatic potential surfaces for the $FeMg_3^+$ isomers before and after $H_2$ adsorption (Figure A6 in SF) are similar to those of $NiMg_3$ isomers as discussed in section 3.2.1.2.  Moreover, analyses of the other electronic properties such as the molecular orbitals and the charge transfer processes in $FeMg_3^+$ isomers before and after $H_2$ adsorption also follows the explanations of those in $NiMg_3$ isomers.  The molecular orbitals and electronic configurations of bare and hydrogen adsorbed $FeMg_3^+$ isomers are provided in Figure A7 and Table A8 in the SF.  On the basis of these various parameters, it is confirmed that the cluster geometry irrespective of its charge state can dictate the nature of hydrogen adsorption and thereby govern the gravimetric density of the cluster for storage of molecular hydrogen.
\clearpage
\begin{figure}
\centering
\includegraphics[width=0.9\textwidth]{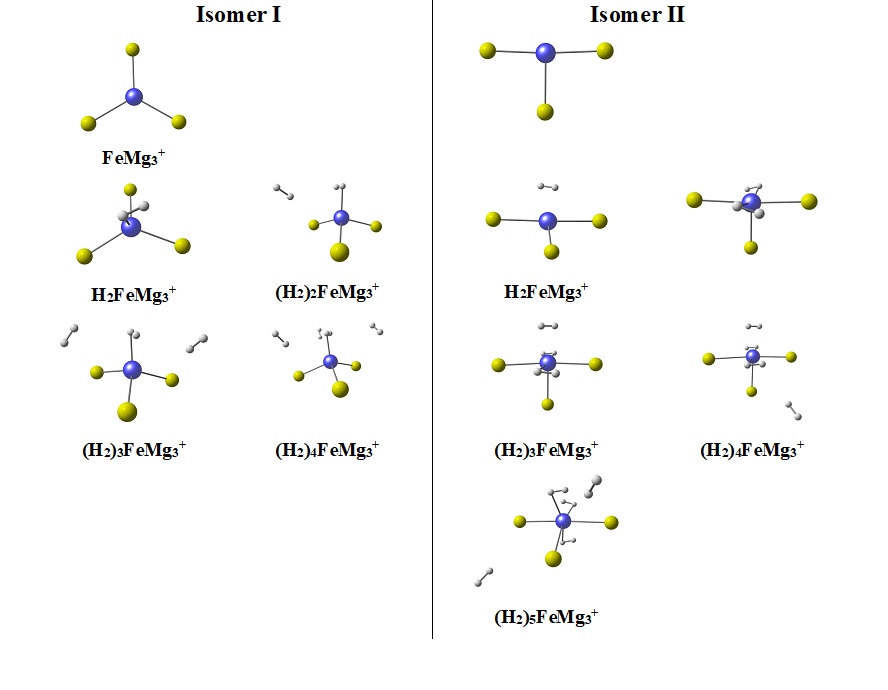}
\caption{\bf Multiple $\bf H_2$ adsorbed complexes of $\bf FeMg_3^+$ isomers.}\label{fig8}
\end{figure}

\begin{figure}
\centering
\includegraphics[width=0.7\textwidth]{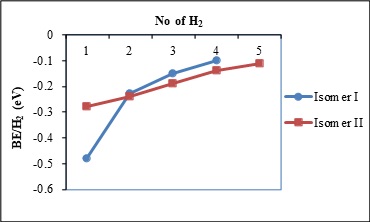}
\caption{\bf Variation of binding energy per $\bf H_2$ molecule (BE/$\bf H_2$) with number of $\bf H_2$ on Isomer I and isomer II of $\bf FeMg_3^+$ cluster.}\label{fig9}
\end{figure}
\clearpage
\section {Conclusion}
We have investigated the molecular hydrogen adsorption properties of $TMMg_3^{0,+}$ clusters.  It has been found that doping of 3d TM atoms into $Mg_4^{0,+}$ clusters convert the molecular hydrogen adsorption process in the clusters from endothermic to exothermic nature.   Hydrogen binding with $TMMg_3^{0,+}$ clusters is dependent on the TM(3d) orbital contributions of the clusters to their frontier MOs.  Molecular $H_2$ is adsorbed in $TMMg_3^{0,+}$ clusters through donor acceptor interaction.  It has also been observed that shifting of HOMO positions towards lower energy in $TMMg_3^+$ clusters increase their TM(3d) orbital contributions in the HOMOs.  This leads to enhancement of $H_2$ binding energies in cationic $TMMg_3$ clusters from their neutral counterparts.  That is why exothermic $H_2$ adsorption takes place in all $TMMg_3^+$ clusters in contrast with only $VMg_3$ and $NiMg_3$ among the neutral clusters.  Adsorption energy for single $H_2$ is found to be the maximum for$ NiMg_3$ (-0.76 eV) and $FeMg_3^+$ (-0.48 eV) clusters.  Based on these results, we can suggest TM(3d) orbitals participation in the HOMOs of the host $TMMg_3^{0,+}$ clusters as the necessary criterion for molecular hydrogen adsorption and this property can be tuned by the dopant as well as the charge state of the cluster.  Further, the structure of the $TMMg_3^{0,+}$ clusters determines the adsorption site for the first adsorbed $H_2$ molecule, i.e., the geometry of the $H_2TMMg_3^{0,+}$ adsorption complex.  This is crucial for step-wise addition of further $H_2$ molecules to the $H_2TMMg_3^{0,+}$ complex.  A single $H_2$ molecule interacts with $TMMg_3^{0,+}$ clusters mainly through the TM atoms via charge transfer mechanism as observed from the density of states and donor-acceptor analyses of our computed data.  Studies of adsorption complexes with multiple hydrogen molecules signify that all the $H_2$ compete to interact with the TM atoms/ions of the clusters.  Therefore, it is also required for the $TM^{0,+}$ to maintain peripheral positions in the clusters so that they can get exposed to all the $H_2$ molecules.  Apart from this, moderate interaction instead of strong binding with the first $H_2$ molecule, in fact, can sustain for larger number of hydrogen adsorption in $TMMg_3^{0,+}$ clusters.  High symmetric clusters with the $TM^{0,+}$ dopants with higher Mg coordination support strong interaction with the first $H_2$ molecule and hence get saturated with less number of such molecules.  In such clusters, all the coordinated Mg atoms help the $TM^{0,+}$ indirectly to interact with the first $H_2$ molecule, which become feeble soon in the successive steps.  The underlying factors to the strength of cluster-hydrogen interaction are the molecular electrostatic potential surfaces; orientation of molecular orbitals of the cluster with those of $H_2$; donation and back-donation of electronic charges between the cluster and $H_2$ molecules.  Among the clusters considered in our study, two higher energy low symmetrical isomers of $NiMg_3$ and $FeMg_3^+$ clusters are found to adsorb larger number of $H_2$ molecules than the corresponding lowest energy symmetric ground states.  The gravimetric densities of these two isomers are 13.28 wt\% and 7.89 wt\%, respectively.  This study, in near future, will open up new directions in the investigation of suitable hydrogen storage materials for practical applications.

\paragraph {Acknowledgement}
BB thanks University Grants Commission, India for research fellowship.

\end{document}